\documentclass[12pt]{article}

\usepackage{hyperref}
\hypersetup{colorlinks,linkcolor={blue},citecolor={red},urlcolor={red}} 
\usepackage{scicite}
\usepackage{tikz,xcolor,hyperref}
\usepackage{graphicx}
\usepackage{amsmath}
\usepackage{amssymb}
\usepackage{float}
\usepackage[all]{xy}
\usepackage{amsfonts}
\usepackage{dcolumn}
\usepackage{bm}
\usepackage{xcolor}
\usepackage{caption}
\usepackage{hyperref}
\usepackage{multirow}
\usepackage[utf8x]{inputenc}
\usepackage{epstopdf}
\usepackage{capt-of}
\usepackage{dcolumn}   
\usepackage{bm}
\usepackage{dsfont}
\usepackage{relsize}
\usepackage{mathtools}
\definecolor{lime}{HTML}{A6CE39}
\DeclareRobustCommand{\orcidicon}{%
	\begin{tikzpicture}
	\draw[lime, fill=lime] (0,0) 
	circle [radius=0.20] 
	node[white] {{\fontfamily{qag}\selectfont \tiny ID}};
	\draw[white, fill=white] (-0.0625,0.095) 
	circle [radius=0.012];
	\end{tikzpicture}
	\hspace{-2mm}
}

\foreach \x in {A, ..., Z}{%
	\expandafter\xdef\csname orcid\x\endcsname{\noexpand\href{https://orcid.org/\csname orcidauthor\x\endcsname}{\noexpand\orcidicon}}
}

\usepackage{times}

\newenvironment{sciabstract}{%
\begin{quote} \bf}
{\end{quote}}

\newcounter{lastnote}

\title{Gravitational baryogenesis in extended teleparallel theories of gravity}

\author
{Snehasish Bhattacharjee \orcidA{}\\
\\
\normalsize{Department of Astronomy, Osmania University, Hyderabad, India, 500007}\\
\normalsize{snehasish.bhattacharjee.666@gmail.com}\\
\normalsize{}\\
\\
}

\date{\today}

\begin{document} 


\baselineskip24pt


\maketitle


\begin{sciabstract}
  The article communicates gravitational baryogenesis in non-minimal $f(T)$ gravity and $f(T,B)$ teleparallel gravity where $T$ denote the torsion scalar and $B$ a boundary term. These extended teleparallel theories of gravity differ from the usual $f(T)$ gravity and therefore could provide key insights in expounding the baryon asymmetry of the universe. Furthermore, such a study also put constraints on the model parameters of these extended teleparallel theory of gravity. I present different baryogenesis interactions proportional to $\partial_{i}T$, $\partial_{i}f(T)$, $\partial_{i}(T+B)$ and $\partial_{i}f(T+B)$ and find that both of these teleparallel theories of gravity yield viable estimates of the baron-to-entropy ratio compatible with observations except for the baryogenesis interaction proportional to  $\partial_{i}(T+B)$. It is therefore encouraging to exercise these extended theories of gravity in other cosmological areas to under their efficiency and applicability in characterizing the current state of the universe.
\end{sciabstract}

\section{Introduction}
Cosmological observations \cite{1to2} revealed that the universe contain negligible amount of anti-matter. It is believed that just moments after the the big-bang, an unknown asymmetry between matter and anti-matter emerged which transformed a tiny fraction of anti-matter into ordinary matter. Matter and anti-matter then annihilated leaving behind the tiny excess matter which accounts for all the matter we see around us. However, the exact nature or cause for this asymmetry is unknown and is termed as "baryon asymmetry". Baryon asymmetry implies there exists some fundamental differences between matter (for instance, electron) and anti-matter (i.e, positron). Nevertheless, many well motivated cosmological theories have lined up to explain this conundrum \cite{3to15}. In these theories, exotic interactions not limited to the standard model have been proposed to account for this dissimilarity. \\
Gravitational baryogenesis is a very promising cosmological theory put forward in \cite{16} which employs the Sakharov conditions \cite{23} and assure a baryon asymmetry through the CP-violating interaction defined as 
\begin{equation}\label{eq1}
\frac{1}{M_{D}}\int d^{4}x\sqrt{-g}(\partial_{i}R)J^{i},
\end{equation}
where $M_{D}$ represents the mass of the effective theory, $g$ denote the metric scalar, $J_{i}$ the baryonic current and $R$ the Ricci scalar. Eq. \ref{eq1} was further extended to many modified gravity theories \cite{17to22,barft,bhatta,bhatta2,barfg}. The idea behind the extension of this interaction to modified gravity theories is driven by the fact that other curvature invariants such as Torsion scalar $T$, Gauss-Bonnet scalar $G$, Non-metricity $Q$ provide a non-zero baryon asymmetry in a radiation dominated universe $\omega=1/3$ which cannot be attained in GR. \\
In this paper, I shall investigate the phenomenon of gravitational baryogenesis in the framework of extended teleparallel theories of gravity. To be precise, I shall work with non-minimal $f(T)$ teleparallel gravity and $f(T,B)$ teleparallel gravity where $B$ is a boundary term. \\
In \cite{harko} the non-minimal $f(T)$ teleparallel gravity was introduced which incorporates a non-minimal matter-torsion coupling with a coupling constant $\lambda$. The field equations therefore consists of two arbitrary functions of $T$, namely $f_{1}(T)$ and $f_{2}(T)$ with $f_{2}(T)$ linearly coupled to the matter Lagrangian \cite{harko}. For $\lambda=0$, the field equations reduces to the usual $f(T)$ teleparallel gravity. Constraints on power law, the exponential and the square-root exponential $f(T)$ gravity models from Big Bang Nucleosynthesis have been reported in \cite{sal2}.\\
Another generalization of $f(T)$ teleparallel gravity was introduced in \cite{ftb} by the incorporation of a boundary term $B$ in addition to $T$ in the Lagrangian. By doing so, $f(T,B)$ gravity reduces to both $f(T)$ and $f(R)$ gravity under certain conditions as $R=T+B$. Note that as reported in \cite{sal1}, the functions $f(T)$ and $f(B)$ cannot exist independently and the only choice is $f(T+B)$. In \cite{bhatta4}, the author imposed constraints on power-law, mixed and logarithmic $f(T,B)$ gravity models from the energy conditions. \\
The paper therefore aims at understanding the influence of non-minimal matter-torsion coupling and the boundary term $B$ in explaining the baryon asymmetry of the universe.Furthermore, such a study also put constraints on the model parameters of these extended teleparallel theory of gravity.\\
The paper is organized as follows: In Section \ref{II} and \ref{III} I provide a quick overview of non-minimal $f(T)$ gravity and $f(T,B)$ gravity respectively. in Section \ref{IV} I explain and investigate gravitational baryogenesis in the framework of non-minimal $f(T)$ gravity and $f(T,B)$ gravity and in Section \ref{V} I present the results and conclusions.

\section{Non-Minimal $f(T)$ Gravity}\label{II}
In teleparallel gravity, the dynamical variables are the vierbein fields $e_{A}(x^{\mu})$, which at each point $x^{\mu}$ of the manifold form an orthonormal basis for the tangent space, where $e_{A}.e_{B}=\eta_{AB}$, with $\eta_{AB} = diag (1,-1,-1,-1)$ \cite{harko}. The metric is therefore procured from the dual vierbein as 
\begin{equation}
g_{\mu\nu} (x) = \eta_{AB}e^{A}_{\mu}(x)e^{B}_{\nu}(x).
\end{equation}
Note that in teleparallel gravity the Weitzenbock connection \cite{harko32} replaces the usual Levi-Civita connection appraring in GR and the curvature is replaced by the torion tensor and is defined as \cite{harko}
\begin{equation}
T^{\lambda}_{\mu\nu} =\overset{w \lambda}{\Gamma_{\nu\mu}}-\overset{w \lambda}{\Gamma_{\mu \nu}} = e^{\lambda}_{A}(\partial_{\mu} e^{A}_{\nu}- \partial_{nu}e^{A}_{\mu}).
\end{equation}
The teleparallel Lagrangian also known as torsion scalar can therefore be written as \cite{harko21to24,harko33}
\begin{equation}
T\equiv \frac{1}{2}T^{\rho \mu \nu} T_{\nu \mu \rho} + \frac{1}{4}T^{\rho \mu \nu} T_{\nu \mu \rho} - T_{\rho \mu }{}^{\rho} T^{\nu \mu}{}_{\nu}.
\end{equation}

Now, the action in non-minimal $f(T)$ gravity is given as \cite{harko}
\begin{equation}\label{eq2}
\mathcal{S}=\frac{1}{2} \int d^{4}x e \left[T + f_{1}(T) + (1 + \lambda f_{2}(T)) \mathcal{L}_{m} \right],
\end{equation}
where $f_{1}(T)$ and $f_{2}(T)$ are functions of $T$ and $\lambda$ represents the coupling constant in units of $\text{mass}^{-2}$. variation of the action Eq.\ref{eq2} with respect to the tetrad $e^{A}_{\rho}$ yields the following field equation \cite{harko}
\begin{multline}
( \lambda f^{'}_{2} \mathcal{L}_{m}+ f^{'}_{1} +1) [e^{-1}\partial_{\mu} (ee^{\alpha}_{A}S_{\alpha}^{\rho \mu}) - e^{\alpha}_{A}T^{\mu}{}_{\nu \alpha}S_{\mu}{}^{\nu}\rho] + e^{\rho}_{A} \left(\frac{f_{1}}{4} \right) + (\lambda f^{''}_{2}\mathcal{L}_{m} +  f^{''}_{1}) \partial_{\mu}T e^{\alpha}_{A} S_{\alpha}{}^{\rho \mu} \\ + \lambda f^{'}_{2}e^{\alpha}_{A}S_{\alpha}^{\rho \mu} \partial_{\mu} \mathcal{L}_{m} - \frac{1}{4}\lambda f^{'}_{2} \partial_{\mu} T e ^{\alpha}_{A}\overset{em}{S{}_{\alpha}{}^{\rho \mu}} = 4 \pi G (1 + \lambda f_{2})e_{A}^{\alpha}\overset{em}{T{}_{\alpha}{}^{\rho}},
\end{multline}
where $\overset{em}{S{}_{\alpha}{}^{\rho \mu}} = \frac{\partial \mathcal{L}_{m}}{\partial \partial_{\mu}e^{A}_{\rho}}$ \cite{harko} and prime represents derivatives with respect to $T$. Following \cite{harko35to36}, I set $\mathcal{L}_{m}/ 16 \pi G = -\rho_{m}$ which make $\overset{em}{S{}_{\alpha}{}^{\rho \mu}} =0$. Now, for a perfect fluid and a flat FRW vierbein, the Friedman equations reads \cite{harko}
\begin{equation}\label{eq3}
H^{2} = \frac{8 \pi G}{3} [ \lambda (f_{2} + 12 H^{2}f^{'}_{2})+1]\rho_{m} - \frac{1}{6} (f_{1} + 12 H^{2}f^{'}_{1})
\end{equation}
and 
\begin{equation}\label{eq4}
\dot{H} = - \left[\frac{4 \pi G (p_{m}+\rho_{m}) [ \lambda (f_{2} + 12 H^{2}f^{'}_{2})+1]}{1 - 16 \pi G \lambda \rho_{m}(f^{'}_{2}-12 H^{2}f^{''}_{2}) + f^{'}_{1} - 12 H_{2}f^{''}_{1}}\right].
\end{equation}
Now in a flat FRW universe $T=-6 H^{2}$ and therefore following the convention of \cite{harko}, I can express $f_{1}(T) = f_{1}(H)$ and $f_{2}(T) = f_{2}(H)$. Therefore from Eqs. \ref{eq3} and \ref{eq4}, the general expression of energy density $\rho_{m}$ reads \cite{harko}
\begin{equation}\label{eq5}
\rho_{m} = \left[\frac{ 3 H^{2} + [12 H^{2} f^{'}_{1}(T) +f_{1}(H)]/2}{ 1 + \lambda [12 H^{2} f^{'}_{2}(T) +f_{2}(H)]}\right]. 
\end{equation}
\section{$f(T,B)$ Gravity}\label{III}
In this section I shall present a quick overview of $f(T,B)$ teleparallel gravity introduced in \cite{ftb} which generalizes the standard $f(T)$ teleparallel gravity through the inclusion of a boundary term $B$. The action in $f(T,B)$ teleparallel gravity is defined as \cite{ftb}
\begin{equation}\label{13}
\mathcal{S} = \frac{1}{\kappa}\int d^{4}x e f(T,B) + \mathcal{L}_{m},
\end{equation}
where $B$ is the boundary term and is defined as \cite{ftb} $B=\frac{2}{e}\partial_{i}(e T ^{i}) = \bigtriangledown_{i}T^{i}$. The boundary term $B$ generalizes the $f(T)$ gravity to its metric counterpart $f(R)$ gravity since for $f(T,B) = f(T+B) = f (R)$ \cite{no}. \\
Varying the action \ref{13} with respect to the tetrad, the field equation reads
\begin{multline}\label{14}
16 \pi e \Theta^{\lambda}_{a}=  4 e \left[ \partial_{i}f_{B} + \partial_{i}f_{T}  \right] S_{a}^{i \lambda} - 4 e f_{T} T^{\sigma} _{ia}S_{\sigma}^{\lambda i} - e f E^{\lambda}_{a} \\ + e B E ^{\lambda}_{a}f_{B}+2e E ^{\lambda}_{a} \square f_{B} - 2 e E ^{\sigma}_{a} \bigtriangledown^{\lambda}\bigtriangledown_{\sigma}f_{B}  + 4 \partial _{i} (e S_{a}^{i \lambda}) f_{T} . 
\end{multline}
Assuming a flat FRW spacetime with a perfect fluid, the general expression for the matter energy density $\rho_{m}(t)$ reads \cite{ftb}  
\begin{equation}\label{eq15}
\rho(t) = -3 H^{2} (3 f_{B} + 2 f_{T}) - 3 \dot{H}f_{B} + 3 H \dot{f_{B}} + \frac{1}{2} f.
\end{equation}
Note that for a FRW spacetime, $B=6(\dot{H} + 3 H^{2})$ \cite{ftb}.

\section{Gravitational Baryogenesis}\label{IV}
The baryon-to-entropy ratio at the present epoch is observed to be \cite{1to2} 
\begin{equation}\label{eq7}
\frac{\eta_{B}}{s}\simeq 9 \times 10^{-11}.
\end{equation}
Now in order to suffice the observed profuse abundance of matter over antimatter, Sacharov conditions \cite{23} must be satisfied. These conditions state that in order to have a net baryon excess, processes which violate baryon number, charge ($C$) and joint charge-Parity $CP$ interactions and processes outside of thermal equilibrium must have taken place. \\
As the universe expands, temperature $T$ drops and after a certain value (called critical value) $T_{D}$ processes contributing to baryon asymmetry freezes and the resultant baryon to entropy ratio can be mathematically expressed in the framework of gravitational baryogenesis as \cite{16} 
\begin{equation}\label{eq8}
\frac{\eta_{B}}{s}\simeq - \frac{15 g_{b}}{g_{*s}}\frac{\dot{R}}{M_{D}^{2} T_{D}},
\end{equation}
where $g_{b}$ and $g_{*s}$ represent the total number of degrees of freedom for baryons and massless particles respectively. Additionally, by presuming thermal equilibrium, the energy density $\rho_{m}$ is related to temperature $T$ as 
\begin{equation}\label{rho}
\rho_{m}(T) = \frac{\pi^{2}}{30}g_{*s} T^{4}.
\end{equation}
In subsequent sections, I shall present the viability of non-minimal $f(T)$ gravity and $f(T,B)$ gravity in addressing various gravitational baryogenesis interactions. For the analysis, I assume power law expansion where the scale factor $a(t) \sim t^{n}, n>0$ a constant. 
\subsection{Gravitational Baryogenesis in Non-Minimal $f(T)$ Gravity}
\begin{figure*}[h]
\centering
  \includegraphics[width=7.4 cm]{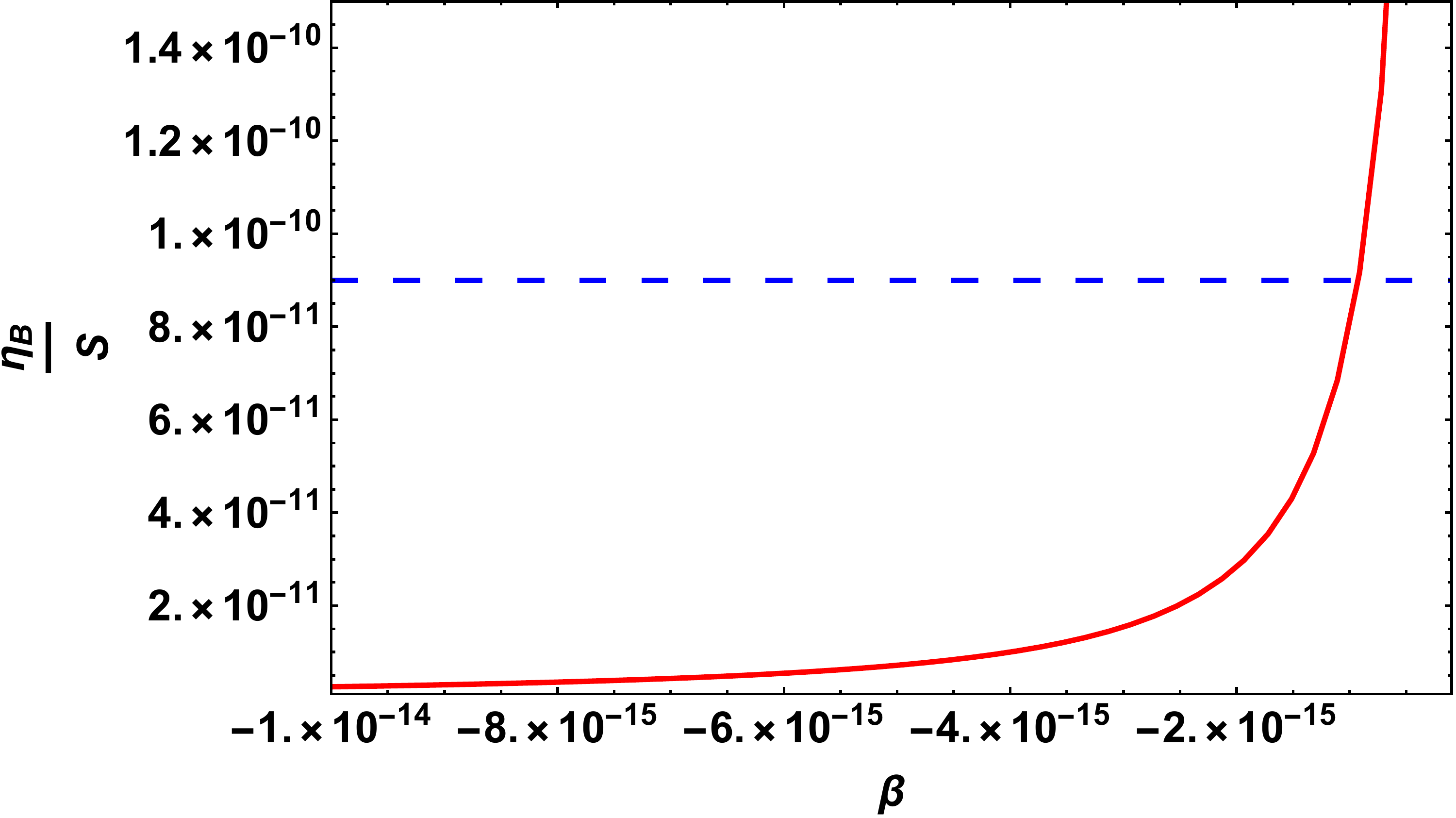}
  \includegraphics[width=7.8 cm]{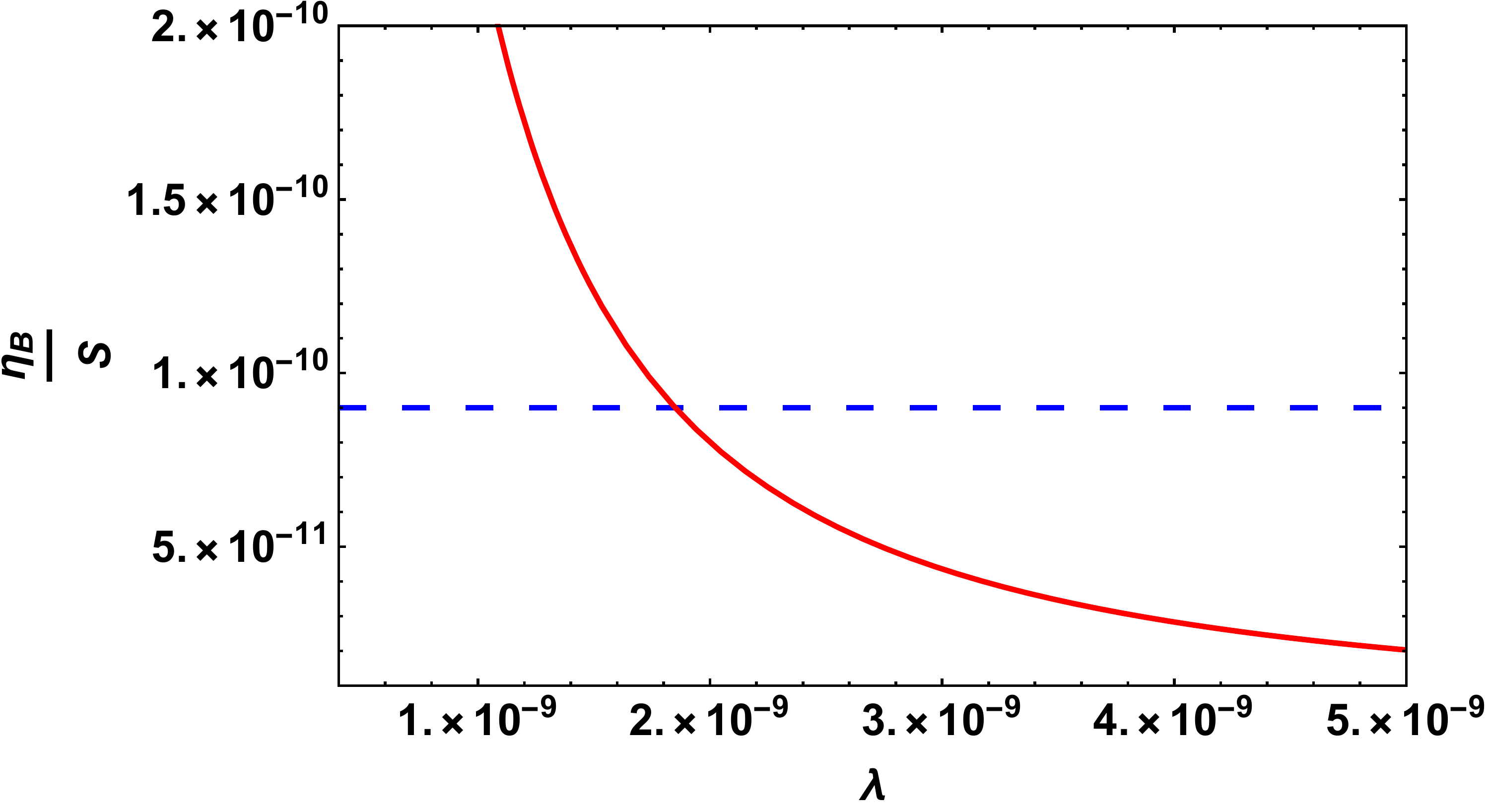}
\caption{Left panel shows $\eta_{B}/s$ as a function of $\beta$. Right panel shows $\eta_{B}/s$ as a function of coupling constant $\lambda$. The plots are drawn for $ n=0.2, \Lambda=1, g_{*s}=106, g_{b} = 1$ and $T_{D}=2\times 10^{12} GeV$ and $M_{D}=2\times 10^{16} GeV$. The dashed blue line corresponds to the observational constraint $\frac{\eta_{B}}{s}\simeq 9 \times 10^{-11}$. }
\label{FIG1}
\end{figure*}

In $f(T)$ teleparallel gravity, the CP-violating interactions are proportional to the torsion $T$ instead of $R$ and can be defined as \cite{barft}
\begin{equation}\label{eq9}
\frac{1}{M_{D}}\int d^{4}x\sqrt{-g}(\partial_{i}T)J^{i}.
\end{equation} 
Thus, the net baryon to entropy ratio can be expressed as \cite{barft}
\begin{equation}\label{eq10}
\frac{\eta_{B}}{s}\simeq  \frac{15 g_{b}}{g_{*s}}\frac{\dot{T}}{M_{D}^{2} T_{D}}.
\end{equation}
In \cite{barft}, the authors studied the viability of some minimally coupled $f(T)$ gravity models in addressing the baryon asymmetry by finding corners in parameter spaces for which Eq. \ref{eq10} produces $\eta_{B}/s$ consistent with observational value \ref{eq7}. Since non-minimally coupled $f(T)$ gravity produces field equations which differ from the usual $f(T)$ gravity, it is therefore encouraging to study the consequences of a CP-violating interaction in non-minimal $f(T)$ gravity models. \\
In this work, I shall assume $f_{1}(T)=\Lambda$ and $f_{2}(T)=\beta T$ where $\Lambda \geq 0$ and $\beta$ are constants. Substituting in Eq. \ref{eq5}, the expression of density $\rho_{m}$ reads
\begin{equation}\label{20}
\rho_{m}=\frac{\frac{\Lambda }{2}+\frac{3 n^2}{t^2}}{\frac{13 \beta  \lambda  n^2}{t^2}+1}.
\end{equation}  
The next step involves equating Eq. \ref{20} with Eq. \ref{rho} to find an analytical expression for the decoupling time $t_{D}$ and reads
\begin{equation}\label{21}
t_{D}=\sqrt{\frac{6 n^2-\frac{1378}{15} \pi ^2 \beta  \lambda  n^2 T_{D}^4}{\frac{106 \pi ^2 T_{D}^4}{15}-\Lambda }}.
\end{equation}
Finally, substituting, Eq. \ref{21} in Eq. \ref{eq10}, the expression for the baryon to entropy ratio for  $g_{*s}=106, g_{b} = 1$ reads 
\begin{equation}\label{22}
\frac{\eta_{B}}{s}\simeq \frac{90 n^2}{53 M_{D}^2 T_{D} \left(\frac{6 n^2-\frac{1378}{15} \pi ^2 \beta  \lambda  n^2 T_{D}^4}{\frac{106 \pi ^2 T_{D}^4}{15}-\Lambda }\right)^{3/2}}.
\end{equation}
Substituting $ n=0.2, \Lambda=1, \beta = -1 \times 10^{-14}, \lambda = 2 \times 10^{-9}, T_{D}=2\times 10^{12} GeV$ and $M_{D}=2\times 10^{16} GeV$, the baryon to entropy ratio reads $\frac{\eta_{B}}{s}\simeq 8 \times 10^{-11}$ which is consistent with observations. In Fig. \ref{FIG1} I show the variation of $\frac{\eta_{B}}{s}$ (Eq. \ref{22}) against the coupling constant $\lambda$ and model parameter $\beta$. 

\subsection{Generalized Baryogenesis Interaction in Non-Minimal $f(T)$ Gravity}
\begin{figure*}[h]
\centering
  \includegraphics[width=7.8 cm]{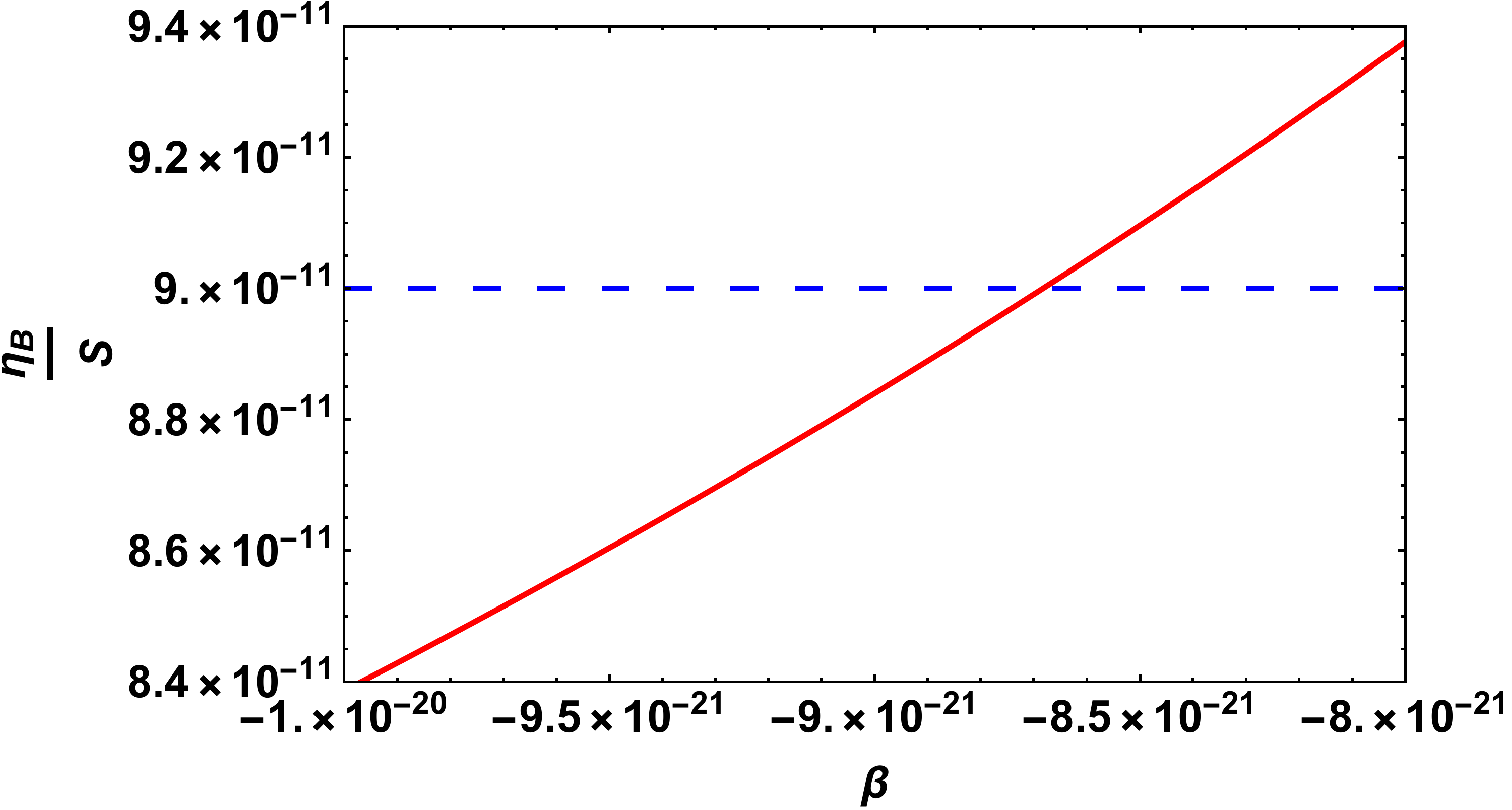}
  \includegraphics[width=7.6 cm]{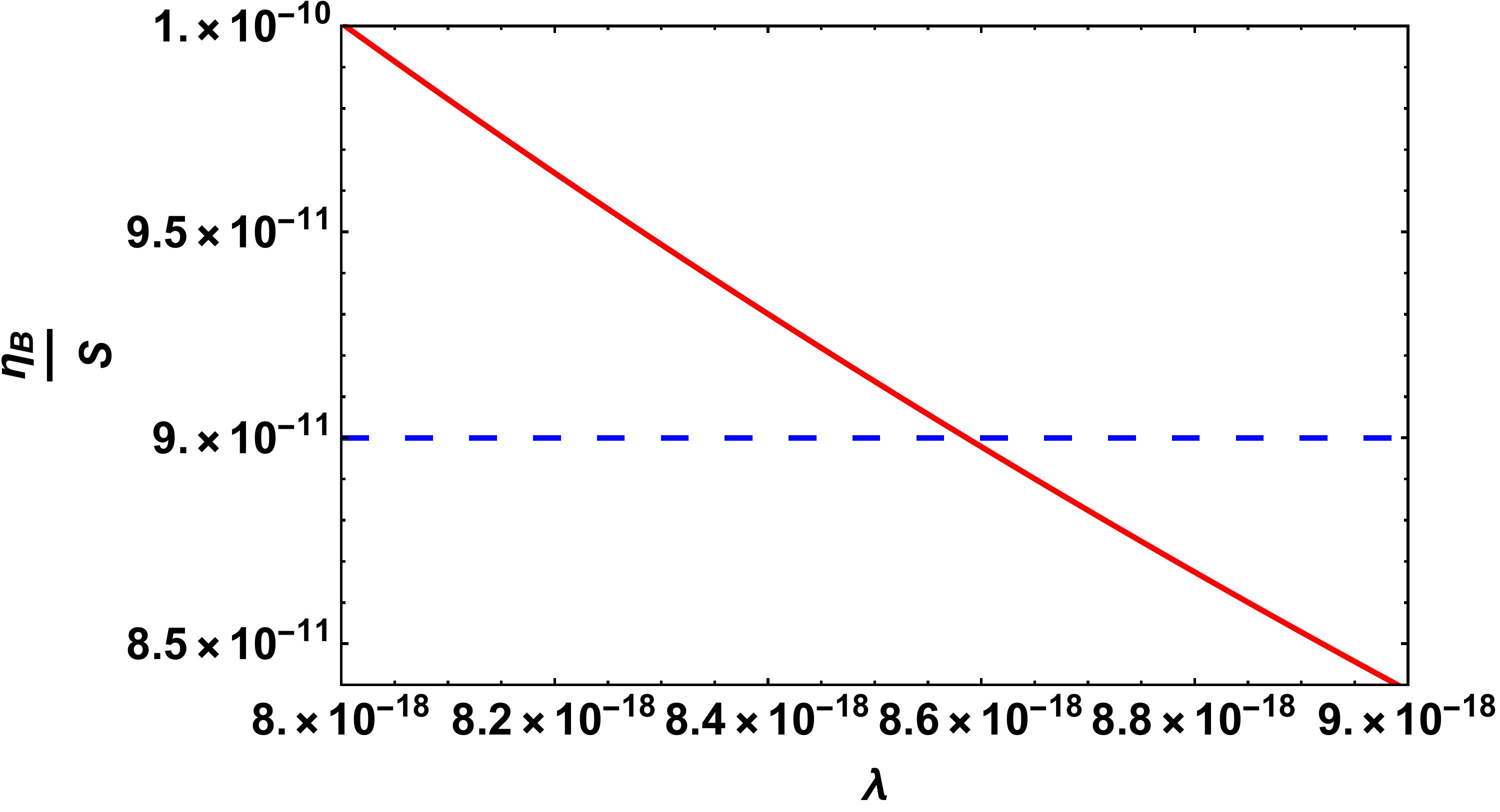}
\caption{Left panel shows $\eta_{B}/s$ as a function of $\beta$. Right panel shows $\eta_{B}/s$ as a function of coupling constant $\lambda$. The plots are drawn for $ n=0.2, \Lambda=1, g_{*s}=106, g_{b} = 1$ and $T_{D}=2\times 10^{12} GeV$ and $M_{D}=2\times 10^{16} GeV$. The dashed blue line corresponds to the observational constraint $\frac{\eta_{B}}{s}\simeq 9 \times 10^{-11}$. }
\label{FIG2}
\end{figure*}

In \cite{barft}, the authors defined a more complete baryogenesis interaction which is proportional to $\partial_{i}f(T)$ instead of $\partial_{i}T$. However, the interaction gets modified for non-minimal $f(T)$ gravity owing to the presence of two functional forms of the torsion $T$ (i.e, $f_{1}(T)$ and $f_{2}(T)$). The CP-violating interaction for non-minimal $f(T)$ gravity can be written as 
\begin{equation}\label{eq11}
\frac{1}{M_{D}}\int d^{4}x\sqrt{-g}(\partial_{i}(f_{1}(T)+f_{2}(T))J^{i}.
\end{equation}
The mathematical form of the baryon-to-entropy ratio corresponding to Eq. \ref{eq11} for $g_{*s}=106, g_{b} = 1$ reads 
\begin{equation}\label{eq12}
\frac{\eta_{B}}{s}\simeq - \frac{15 g_{b}}{g_{*s}}\frac{\dot{T}(f_{1,T}+f_{2,T})}{M_{D}^{2} T_{D}},
\end{equation}
where $f_{1,T}$ and $f_{2,T}$ are derivatives of the functions $f_{1}(T)$ and $f_{2}(T)$ with respect to the torsion $T$ respectively.\\
Now for the relevant model employed in the work, $f_{1,T}=0$ and $f_{2,T}=\beta$. Therefore, the baryon to entropy ratio (Eq. \ref{eq12}) reduces to
\begin{equation}\label{23}
\frac{\eta_{B}}{s}\simeq -\frac{90 \beta  n^2}{53 M_{D}^2 T_{D} \left(\frac{6 n^2-\frac{1378}{15} \pi ^2 \beta  \lambda  n^2 T_{D}^4}{\frac{106 \pi ^2 T_{D}^4}{15}-\Lambda }\right)^{3/2}}.
\end{equation} 
Substituting $ n=0.2, \Lambda=1, \beta = -1 \times 10^{-19}, \lambda = 9 \times 10^{-18}, T_{D}=2\times 10^{12} GeV$ and $M_{D}=2\times 10^{16} GeV$, the baryon to entropy ratio reads $\frac{\eta_{B}}{s}\simeq 8.4 \times 10^{-11}$ which similar to the previous case agrees well with observations. In Fig. \ref{FIG2} I show the variation of $\frac{\eta_{B}}{s}$ (Eq. \ref{23}) against the coupling constant $\lambda$ and model parameter $\beta$. From the analysis it is clear that very small values for the coupling constant $\lambda$ and the model parameter $\beta$ are required to suffice the baryon asymmetry of the universe. Interestingly, this corresponds to a minute deviation from general relativity since for $\lambda = \beta =0$ the field equations correspond to the teleparallel equivalent of general relativity. 
\subsection{Gravitational Baryogenesis in $f(T,B)$ Gravity}
In $f(T,B)$ teleparallel gravity, the baryogenesis interactions are proportional to the boundary term $B$ in addition to the torsion scalar $T$. Therefore, the CP-violating interaction can be expressed as 
\begin{equation}\label{eq16}
\frac{1}{M_{D}}\int d^{4}x\sqrt{-g}(\partial_{i}(T+B))J^{i}.
\end{equation}
The expression for the baryon to entropy ratio therefore reads 
\begin{equation}\label{eq17}
\frac{\eta_{B}}{s}\simeq  -\frac{15 g_{b}}{g_{*s}}\frac{(\dot{T}+\dot{B})}{M_{D}^{2} T_{D}}.
\end{equation}
In this work, I assume the functional form $f(T,B)=\alpha  (T B)^m$ where $\alpha, m$ are constants. Substituting this in the expression of density $\rho_{m}$ (Eq. \ref{eq15}), I obtain  
\begin{equation}\label{24}
\rho_{m} = \frac{\alpha  36^m (m (-0.67 m+0.5 n+0.17)+0.5 n-0.17) \left(\frac{(1-3 n) n^3}{t^4}\right)^m}{ n-0.33}.
\end{equation}
Equating Eq. \ref{rho} with Eq. \ref{24}, the expression for the decoupling time $t_{D}$ reads
\begin{equation}\label{26}
t_{D}=\left(\left(\frac{5}{53}\right)^{1/m} \pi ^{-2/m} (1-3 n) n^3 \left(\frac{3^{-2 m-1} 4^{-m} (n-0.33)T_{D}^4}{\alpha  (m (-0.67 m+0.5 n+0.17)+0.5 n-0.17)}\right)^{-1/m}\right)^{0.25}.
\end{equation}
Finally, substituting all the respective values in Eq. \ref{eq17}, the baryon to entropy ratio reads
\begin{equation}\label{25}
\frac{\eta_{B}}{s}\simeq -\frac{180 n^2}{53 M_{D}^2 T_{D} \left(\left(\frac{5}{53}\right)^{1/m} \pi ^{-2/m} (1-3 n) n^3 \left(\frac{3^{-2 m-1} 4^{-m} (n-0.33) T_{D}^4}{\alpha  (m (-0.67 m+0.5 n+0.17)+0.5 n-0.17)}\right)^{-1/m}\right)^{0.75}}.
\end{equation}
Unfortunately, for this type of CP-violating interaction, a viable estimate of the baryon to entropy ratio cannot be obtained. Nonetheless, in the next section I shall show that a observationally congruous $\frac{\eta_{B}}{s}$ is possible if the CP-violating interactions are made proportional to $\partial_{i}f(T+B)$ rather than $\partial_{i}(T+B)$.
\subsection{Generalized Baryogenesis Interaction in $f(T,B)$ Gravity}
\begin{figure*}[h]
\centering
  \includegraphics[width=7.8 cm]{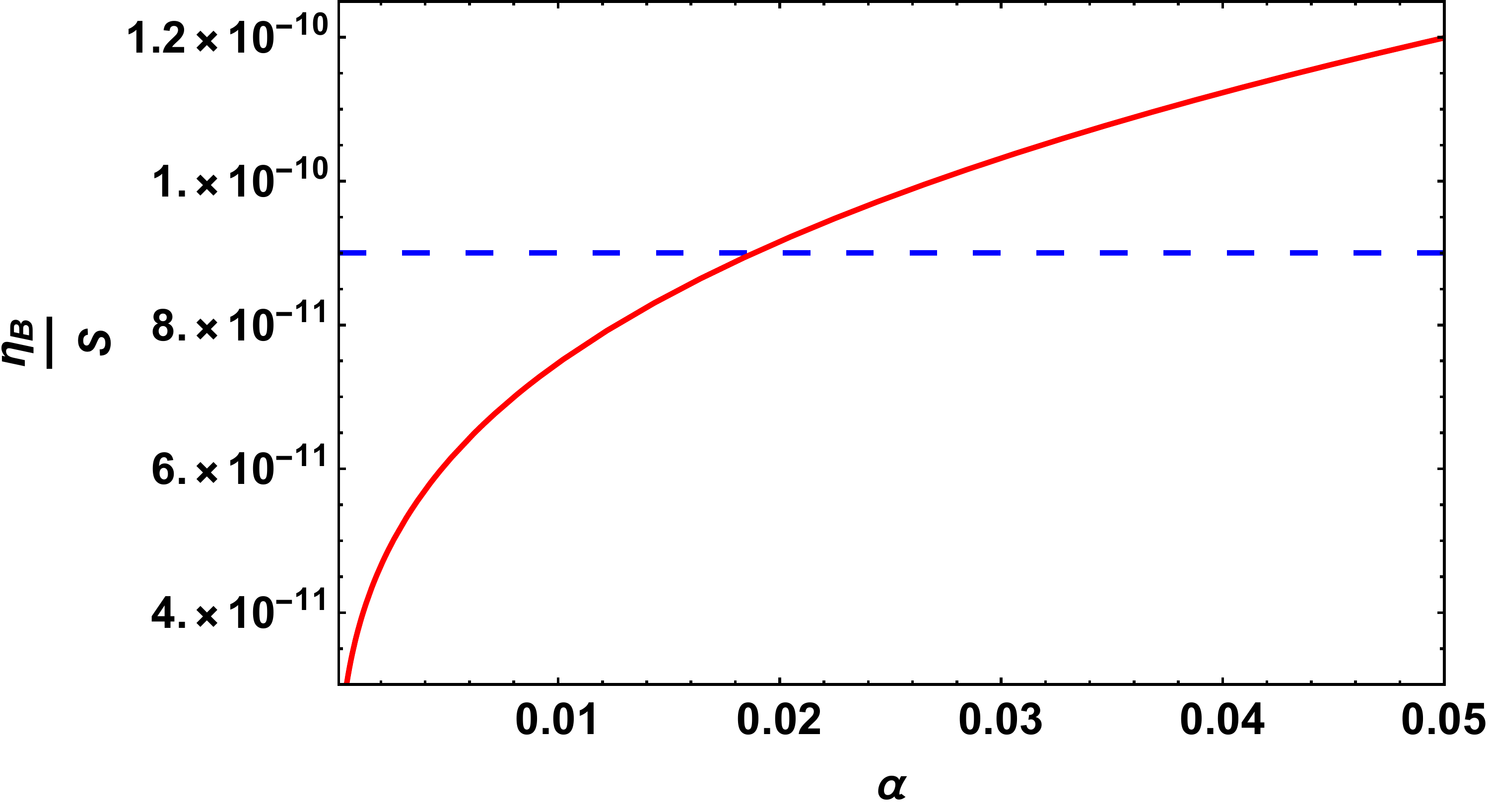}
  \includegraphics[width=7.6 cm]{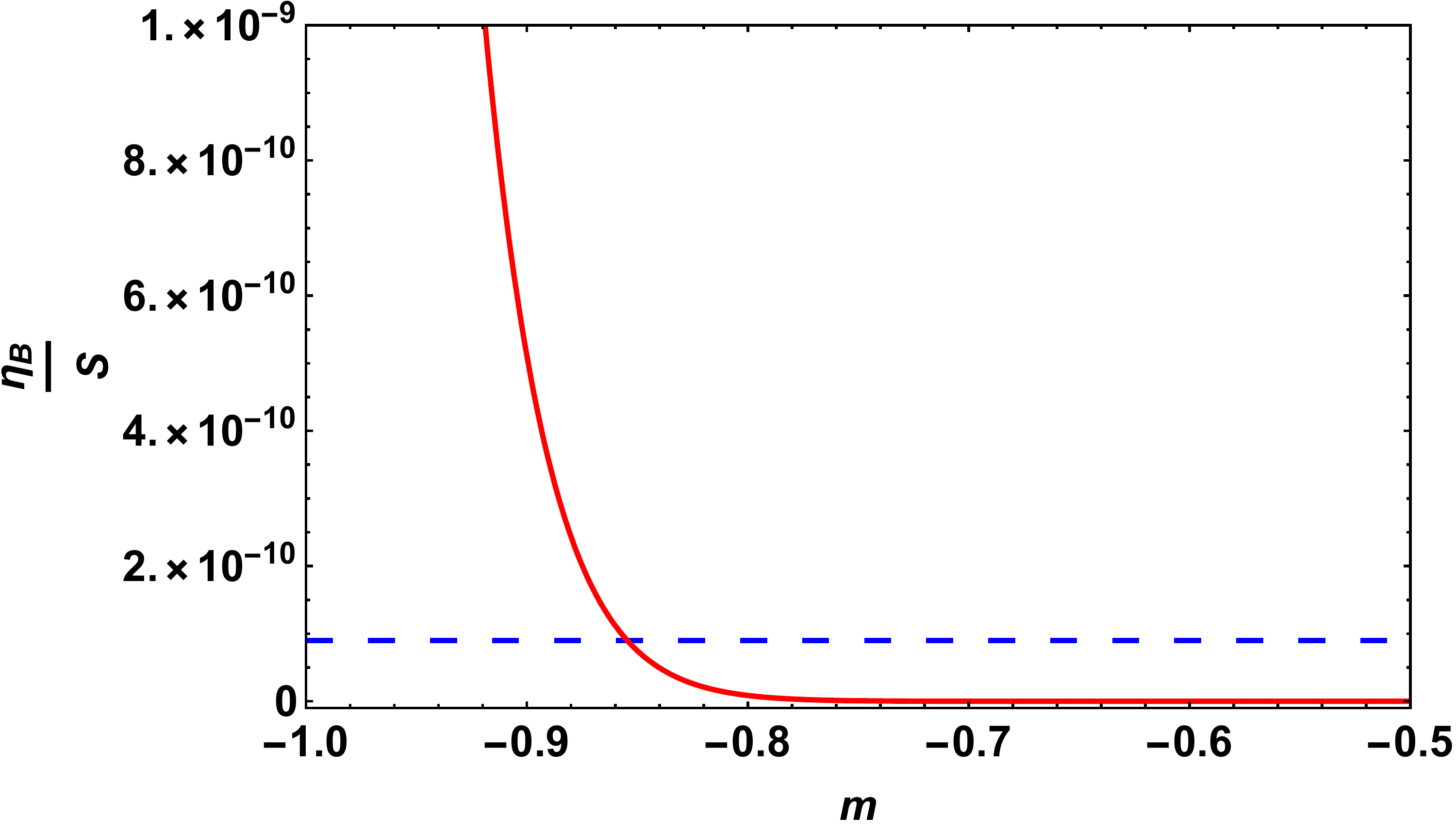}
\caption{Left panel shows $\eta_{B}/s$ as a function of $\alpha$. Right panel shows $\eta_{B}/s$ as a function of $m$. The plots are drawn for $ n=0.3, g_{*s}=106, g_{b} = 1$ and $T_{D}=2\times 10^{12} GeV$ and $M_{D}=2\times 10^{16} GeV$. The dashed blue line corresponds to the observational constraint $\frac{\eta_{B}}{s}\simeq 9 \times 10^{-11}$. }
\label{FIG3}
\end{figure*}

In this section I shall present the generalized baryogenesis interaction for $f(T,B)$ modified gravity in which the CP-violating interactions are proportional to  $\partial_{i}f(T+B)$. The CP-violating interaction can therefore be written as 
\begin{equation}\label{eq18}
\frac{1}{M_{D}}\int d^{4}x\sqrt{-g}(\partial_{i}f(T+B)J^{i}.
\end{equation}
Finally, the baryon to entropy ratio for this type of baryogenesis interaction reads 
\begin{equation}\label{eq19}
\frac{\eta_{B}}{s}\simeq - \frac{15 g_{b}}{g_{*s}}\frac{(\dot{T}f_{T}+\dot{B}f_{B})}{M_{D}^{2} T_{D}}.
\end{equation}
I proceed to substitute Eq. \ref{26} in Eq. \ref{eq19} to obtain
\begin{equation}\label{99}
\frac{\eta_{B}}{s}\simeq \frac{1.7 \alpha  36^m m (1. n-0.17) \left(7^{1/m} \pi ^{2/m} \left(\frac{6^{-2 m} (n-0.33) T_{D}^4}{\alpha  \left(-1.33 m^2+1. m n+0.33 m+1. n-0.33\right)}\right)^{1/m}\right)^m}{M_{D}^2 (3 n-1) T_{D} \left(-0.14^{1/m} \pi ^{-2/m} n^3 (3 n-1) \left(\frac{6^{-2 m} (n-0.33) T_{D}^4}{\alpha  \left(-1.33 m^2+1. m n+0.33 m+1. n-0.33\right)}\right)^{-1/m}\right)^{0.25}}.
\end{equation}
Substituting $ n=0.3, m=-0.85,  \alpha = 1 \times 10^{-2}, T_{D}=2\times 10^{12} GeV$ and $M_{D}=2\times 10^{16} GeV$, the baryon to entropy ratio reads $\frac{\eta_{B}}{s}\simeq 7.4 \times 10^{-11}$ which is very close to the observational constraint. In Fig. \ref{FIG3} I show the variation of $\frac{\eta_{B}}{s}$ (Eq. \ref{99}) against the model parameters $\alpha$ and $m$.
\section{Conclusions}\label{V}
In this paper I investigated the phenomenon of gravitational baryogenesis in the framework of non-minimal $f(T)$ gravity and $f(T,B)$ teleparallel gravity. In these extended teleparallel theories of gravity, the field equations differ from the standard $f(T)$ gravity and is therefore encouraging to study the influence of non-minimal matter-torsion coupling and the boundary term $B$ in expounding the baryon asymmetry of the universe. Furthermore, such a study also put constraints on the model parameters of these extended teleparallel theory of gravity.\\
I first investigate the role of non-minimal torsion coupling for the model $f_{1}(T)=\Lambda$ and $f_{2}(T)=\beta T$ where $\Lambda \geq 0$ and $\beta$ are constants in producing a net $\frac{\eta_{B}}{s}$ consistent with observations for the CP-violating interactions proportional to $\partial_{i}T$ and $\partial_{i}(f_{1}(T)+f_{2}(T))$. For both cases, an observationally acceptable $\frac{\eta_{B}}{s}$ is obtained which significantly increases the efficiency and applicability of non-minimal $f(T)$ gravity in other cosmological areas.\\
I then proceed to study the role of the boundary term $B$ in addition to the torsion $T$ for the model $f(T,B)=\alpha  (T B)^m$ where $\alpha, m$ are constants in addressing the baryon asymmetry of the universe. In this class of modified gravity, the CP-violating interactions are proportional to $\partial_{i}(T+B)$ and $\partial_{i}f(T+B)$. Unfortunately, for the CP-violating interaction proportional to $\partial_{i}(T+B)$, there exists no parameter space for which an acceptable $\frac{\eta_{B}}{s}$ is obtained. Nonetheless, if the CP-violating interactions are made proportional to $\partial_{i}f(T+B)$, an acceptable baryon-to-entropy ratio is obtained for the relevant model employed in the work. \\
The present study also put strict constraints on the parameter spaces for both the teleparallel gravity models. Interestingly, the present study also implies that a net baryon asymmetry seems inevitable if the cosmic evolution were to be governed by teleparallel gravity instead of GR. As a final note, I add that the limits obtained from the present study should be complemented from other cosmological observations coupled with robust theoretical predictions to understand the cosmological applicability of these extended teleparallel theories of gravity.  
\section*{Acknowledgments}
I thank the referees for their suggestions.

\end{document}